\documentclass[namedreferences]{solarphysics}

\usepackage[hyperref,optionalrh]{spr-sola-addons} 
\usepackage{graphicx}        
\usepackage{color}           
\usepackage{breakurl}        
\renewcommand{\vec}[1]{{\mathbfit #1}}

\renewcommand{\div}{ \mathrm{div} }
\newcommand{\cur}{ \mathrm{curl} }

\newcommand{\pder}[2]{ \frac{\partial #1}{\partial #2} }
\newcommand{\grad}{ {\bf \nabla } }

\chardef\us=`\_

\begin{document}

\begin{article}
\begin{opening}

\title{Thermal Trigger for Solar Flares I: 
Fragmentation of the Preflare Current Layer}

\author[addressref={1},corref,email={leonid.ledentsov@gmail.com}]{\inits{L.S.}\fnm{Leonid}~\lnm{Ledentsov}
\orcid{0000-0002-2701-8871}}
\address[id={1}]{Sternberg Astronomical Institute, 
Moscow State University, 
Moscow 119234, Universitetsky pr., 13, Russia}

\runningauthor{Ledentsov L.S.}
\runningtitle{Thermal Trigger for Solar Flares I}

\begin{abstract}
We consider the effects of heat balance on the structural stability of a preflare current layer. 
The problem of small perturbations is solved in the piecewise homogeneous magnetohydrodynamic (MHD) approximation 
taking into account viscosity, electrical and thermal conductivity, 
and radiative cooling.
Solution to the problem allows the formation of an instability of thermal nature. 
There is no external magnetic field inside the current layer in equilibrium state, 
but it can penetrate inside when the current layer is disturbed.
Formation of a magnetic field perturbation inside the layer creates 
a dedicated frequency in a broadband disturbance subject to thermal instability.
In the linear phase, 
the growth time of the instability is proportional to the characteristic time of radiative cooling of the plasma 
and depends on the logarithmic derivatives of the radiative cooling function 
with respect to the plasma parameters.
The instability results in transverse fragmentation of the current layer 
with a spatial period of $1-10 {\rm \, Mm}$
along the layer 
in a wide range of coronal plasma parameters. 
The role of that instability in the triggering of the primary energy release in solar flares is discussed. 
\end{abstract}
\keywords{Plasma Physics; Magnetohydrodynamics; Magnetic Reconnection, Theory; Instabilities; Flares, Models}
\end{opening}
\newpage
\section{Introduction}
     \label{sec1} 

In recent decades, space observatories have made it possible 
to study the development of solar flares in all the ranges of the electromagnetic radiation \citep{2017LRSP...14....2B}. 
The brightness of flare coronal loops in the ultraviolet range is 
one of the most spectacular manifestations of a solar flare
which has been observed in detail. 
The complex structure of the distribution of bright loops in space indicates 
the heterogeneity of the primary energy release in a flare 
\citep{2003ApJ...595L.103K, 2015SoPh..290.2909R}. 
Nevertheless, quasiperiodicity in the spatial distribution of bright loops in a flare arcade can often be noticed.
The Bastille day flare is a telling example of a well-observed flare arcade 
extending over the photospheric neutral line \citep{2000ApJ...540.1126A, 2002ApJ...579..863S}.

According to current understanding, a thin current layer is formed 
over the arcade of magnetic loops before the flare 
\citep{2002A&ARv..10..313P, 2006ASSL..341.....S, 2019LRSP...16....3T}. 
This current layer separates colliding magnetic fluxes 
preventing them to reconnect. 
This leads to the accumulation of free energy 
in a non-potential magnetic field associated with the current. 
Free energy is released in the form of a solar flare 
during fast magnetic reconnection 
when the preflare current layer is destroyed 
\citep{1998A&A...331.1078O, 2000A&A...354..703S, 2007ApJ...671.2139U}. 
The aim of this work is to search for a mechanism 
that can lead to the destruction of the current layer that is quasiperiodic in space.

The effect of the decay of the current layer into individual 
current filaments is known as tearing instability \citep{1963PhFl....6..459F, 1993SSRv...65..253S}. 
This process separates the current layer along streamlines 
facilitating the transition from slow reconnection to fast one. 
However, it does not allow to see in which places along the current direction 
one should expect an increased energy release. 
The current layer decays entirely in the classical tearing instability. 
From the mathematical point of view, this is due to the 
absence of a wave-type solution in the direction along the current. 
Often, a similar solution was sought in the interaction of the current layer 
with magnetohydrodynamic (MHD) waves 
\citep{1976ApJ...205..868V, 2006A&A...452..343N, 2012SoPh..277..283A}. 
Also, a spatially inhomogeneous energy release was considered 
as a result of the corrugation instability of a coronal arcade \citep{2017SoPh..292..184K}. 
The magnetic field frozen into the plasma displaced by the instability could reconnect 
with the overlying magnetic field, 
leading to the heating of the unstable flux tube.

In \cite{1982SoPh...75..237S}, 
the heat balance inside the current layer \citep{1976SvAL....2...13S}
is considered. 
In fact, a particular case 
of thermal instability \citep{1965ApJ...142..531F} in the geometry of the current layer
is studied.    
Investigation of the heat balance in the coronal plasma is applied 
in modeling the observed properties of magnetic loops \citep{2019SoPh..294..173K, 2020PPCF...62a4016A} 
and prominences \citep{2006A&A...460..573C}.
Thermal imbalance leads to the unstable growth of entropy waves \citep{2007AstL...33..309S}
affecting the stability of magnetosonic waves 
\citep{2019A&A...624A..96C, 2020PhPl...27c2110P}
and causing the dispersion of slow MHD waves \citep{2019PhPl...26h2113Z}.
The heat-induced attenuation of slow waves 
in the cylindrical geometry of a magnetic tube \citep{2017ApJ...849...62N}
is used to diagnose the plasma in coronal loops on the Sun \citep{2019A&A...628A.133K}.

We consider a piecewise homogeneous model of a current layer, 
which consists of a magnetically neutral current layer 
surrounded by a plasma with an external magnetic field. 
In the equilibrium state, the plasma inside the current layer does not contain a magnetic field. 
However, the disturbance of the external magnetic field can penetrate inward 
when the screening currents are disturbed.
The situation of the appearance of a magnetic field in an MHD medium 
that does not initially contain one is realized. 
This situation is interesting in itself, and not only in the context of magnetic reconnection. 
Therefore, we first consider the more general problem 
of the heat balance of a homogeneous plasma without a magnetic field (Section~\ref{sec2}). 
Then we apply the found solution to the particular geometry of the preflare current layer (Section~\ref{sec3}).
Finally, we consider this current layer in the context of a coronal plasma (Section~\ref{sec4}).
Our conclusions are given in Section~\ref{sec5}.
    
\section{Thermal Instability of a Homogeneous Plasma} 
      \label{sec2}      

In order to study the physical nature of the process of instability formation, 
homogeneous plasma 
in the single-fluid dissipative MHD approximation is considered.
The MHD approximation has been successfully used for coronal applications 
for more than 50 years 
(e.g. \citeauthor{2020ARA&A..58..441N}, \citeyear{2020ARA&A..58..441N}).
It imposes 
some restrictions on the possible plasma processes under consideration.
First, these processes must be sufficiently slow 
compared with the time of electron-ion collisions, 
so that the Maxwell distribution of electrons and ions 
with a common temperature is established in the plasma.
Plasma processes must be  
also sufficiently slow with respect to 
the inverse plasma conductivity to 
neglect the displacement current in comparison with 
conductive current in Maxwell's equations. 
Second, the magnetic field must be weak enough 
to use isotropic conductivity in the generalized Ohm's law. 
Third, the velocities of the considered plasma motions 
must be sufficiently small in comparison with the speed of light 
so that the action of electric forces 
as compared with magnetic ones can be neglected in the nonrelativistic limit.

The first condition satisfies our consideration 
of the preflare state of the plasma in the solar corona, 
when fast energy release does not yet take place, 
and the separation of electron and ion temperatures is not important.
The second condition is consistent with the general idea of a solar flare 
as a result of the process of magnetic reconnection at the zero point of the magnetic field.
The third condition is certainly valid in the context 
of the observed preflare plasma velocities in the solar corona.
However, the effects of finite conductivity 
during the formation of the preflare current layer cannot be neglected.
The subject of this study is the thermal balance 
of the plasma in the preflare configuration, 
and therefore it is assumed that Joule and viscous heating, 
thermal conductivity, and radiative cooling in the energy equation are preserved.
Thus, the following set of dissipative MHD equations is
sufficient for our consideration 
\citep{1958ForPh...6..437S, 2006ASSL..340.....S}:

\vspace{5mm}
\begin{displaymath}
    \frac{\partial n}{\partial t} + \div \, (n \vec{v})
    = 0 \, ,
\end{displaymath}
\begin{displaymath}
    \mu n \, \frac{\rm{d} \vec{v}}{{\rm d} t}
  = - \grad (2 n k_{_{ \rm B }} T)
  - \frac{1}{4\pi} \, ( \vec{B} \times \cur \vec{B} ) + \eta \, \Delta \vec{v} + \nu \, \grad \, \div \, \vec{v} \, ,
\end{displaymath}
\begin{displaymath}
    \frac{2 n k_{_{ \rm B }} }{\gamma -1} \,
    \frac{{\rm d} T}{{\rm d} t}
  - 2 k_{_{ \rm B }} T \, \frac{ {\rm d} n }{ {\rm d} t } \\
  = \frac{ c^2 }{ ( 4 \pi )^2 \sigma } \, ( {\cur} {\vec B} )^2 + \pder{}{r_\alpha} (\sigma_{\alpha \beta} v_\beta)
  + {\div} \, ( \kappa \grad T ) - \lambda \, (n,T) \, ,
  \end{displaymath}
\begin{displaymath}
    \frac{\partial {\vec B}}{\partial t}
  = {\cur} \, ( {\vec v} \times {\vec B} )
  - \frac{ c^2 }{ 4 \pi } \, {\cur}
    \left( \frac{1}{\sigma} \, {\cur} {\vec B} \right) ,
\end{displaymath}
\begin{equation}
    {\div} {\vec B} = 0 \, .
\label{01}
\end{equation}
Here, $ \mu =1.44 \, m_H $, $ m_H $ is the mass of the hydrogen atom,
$ k_{_{ \rm B }} $ is the Boltzmann constant,
$ \gamma $ is the heat capacity ratio,
$ \kappa $ and $ \sigma $ are the thermal and electric conductivities of the plasma, 
$ \lambda\,(n,T) $ is the radiative cooling function,
$ \eta $ and $ \nu $ are viscosity ratios,
and $\sigma_{\alpha \beta}$ is the viscous stress tensor. 
Transfer coefficients are isotropic in the absence of an external magnetic field.
The heat capacity ratio is assumed $ \gamma = 5/3 $ for simplicity.
$T$ is the temperature, $n$ is the palsma density, $v$ is the plasma velocity, and $B$ is the magnetic field. 
System (\ref{01}) will be also used to describe a piecewise homogeneous model 
of the current layer in Section~\ref{sec3}.
%
%

\subsection{Increments of Instability} 
  \label{sec2.1}

The solution to Equations \ref{01}
in the form of the sum of a constant homogeneous term 
and a small perturbation is sought using the following Fourier
transform with subsequent linearization in $f'$
\begin{displaymath}
f ({\vec r},t) = f_{const} + f' \, {\rm exp} \, ( - i \omega t + i (\vec{k} \vec{r}) ) \, .
\end{displaymath}
Here $ f' \equiv \{{\vec{v}'}, n', T', {\vec{B}'}\} $ are perturbation amplitudes.
%
%

Let us set ${\vec{v}}_{\rm const}=0$ and ${\vec{B}}_{\rm const}=0$.
It is worth noting that both Joule and viscous heating 
turn out to be of second order 
in the perturbation and can be neglected in a linear phase.
Only radiative cooling and thermal conductivity affect 
the thermal balance of the plasma in a linear approximation. 
The first seeks to cool the plasma, 
while the second redistributes heat between regions with different temperatures. 
Thus, the plasma tends to cool against the background of small perturbations.
Naturally, this does not contradict the initial heat balance. 
Even if radiative cooling is not compensated 
by Joule or viscous heating in an unperturbed plasma, 
one can consider additional constant heating
as part of the thermal function $\lambda$. 
An additional constant term creates an initial heat balance 
and does not affect small perturbations, since it disappears during the linearization
\citep{1992PPCF...34..411H, 2004A&A...415..705D, 2019A&A...624A..96C}.
It is also possible to consider a more general non-constant thermal function, 
but such a consideration goes beyond the physical formulation of our problem 
\citep{1978ApJ...220..643R, 1993ApJ...415..335I, 2020A&A...644A..33K}.
The set of linear equations will take the following form:
\begin{equation}
\omega \, n' = n \, {(\vec{k}\vec{v}')} \, ,
    \label{02}
\end{equation}
\begin{equation}
i \omega \, n \, {\vec v}'  = i {\vec k} \, \frac{2 k_{_{ \rm B }}}{ \mu } \, ( n T' + T n' ) 
+k^2 \, \eta \, {\vec v}'  + \omega {\vec k} \, \frac{\nu}{n}\,n ' \, ,
    \label{03}
\end{equation}
\begin{equation}
    i \omega \, \frac{ 2 n k_{_{ \rm B }} }{ \gamma - 1 }  \, T' 
  - i \omega \, 2 k_{_{ \rm B }} T \, n' =
 k^2 \, \kappa \, T' +  \pder{ \lambda }{ T } \, T' + \pder{ \lambda }{ n } \, n'  \, ,
    \label{04}
\end{equation}
\begin{equation}
i \omega \, {\vec B}' = \frac{ c^2 }{ 4 \pi \sigma } \, (k^2 \, {\vec B}'-{\vec k} \,({\vec{k}\vec{B}'})) \, ,
    \label{05}
\end{equation}
\begin{equation}
({\vec{k}\vec{B}'})=0 \, .
    \label{06}
\end{equation}
%
%

Equations \ref{02}\,--\,\ref{06} split into two subsystems of equations. 
The perturbations of velocity, concentration, and temperature 
enter only in the first three equations, 
while the perturbation of the magnetic field enters only in Equations \ref{05} and \ref{06}.
In this regard, it is worth paying attention to a couple of nuances. 

First, we assume that $\vec{B}' \ne 0$. 
In this article, we will not investigate the reasons for the occurrence 
of a nonzero magnetic field perturbation in an initially magnetically neutral plasma. 
Such a study goes beyond the framework of our MHD approach and 
requires the use of kinetic theory, such as the Weibel instability \citep{1959PhRvL...2...83W}. 
In this section, we want to show that the formation of a magnetic field perturbation 
creates a dedicated frequency in a broadband disturbance subject to thermal instability. 
In what follows, when considering the piecewise homogeneous model of the preflare current layer (Section~\ref{sec3}), 
we will assume that the perturbation of the magnetic field penetrates 
into the magnetically neutral current layer from the surrounding plasma 
upon dissipation of the screening currents flowing over the surface of the current layer in an unperturbed state. 
The formulation of the problem implies the appearance of a magnetic field 
in a medium that initially does not contain one, 
and this is exactly what we expect in the region of magnetic reconnection. 
External magnetic fields compensate each other inside the current layer in equilibrium state, 
but they can penetrate inside when electric currents are disturbed.

Second, the division of the Equations \ref{01} into subsystems of equations does not indicate 
the formation of several perturbation modes, 
as is the case in a complete MHD system with a magnetic field 
during the formation of entropy, Alfv\'en, and slow and fast magnetoacoustic waves. 
Mode separation occurs when one dispersion relation allows several different solutions, 
but here we have several dispersion relations for one solution. 
It is also not a resonance between different solutions, 
because we are initially looking for one solution that satisfies two conditions. 
The system of Equations \ref{02}\,--\,\ref{04} describes the linear evolution of entropy and sound modes 
in a nonideal hydrodynamic medium, but Equations \ref{05} and \ref{06}
additionally require the occurrence of a magnetic field perturbation.
This disturbance should not be confused with standard fast and slow magnetoacosutic waves 
for which the existence of the initial guiding field is essential.
The system of Equations \ref{02}\,--\,\ref{06} describes a hydrodynamic disturbance that allows a magnetic field to arise.
This type of perturbations requires specific conditions 
that occur in coronal plasma structures such as current layers only.

Two different subsystems allow us to directly determine the frequency of perturbations
which may become unstable according to the scenario described above. 
Let us substitute Equation \ref{06} in Equation \ref{05}
and express the value of $k^2$. 
Then we multiply Equation \ref{03} per the wave vector $\vec k$ 
and replace $(\vec{k}\vec{v}')$ and $k^2$ 
by Equations \ref{02} and \ref{05}, respectively. 
Finally, we exclude one of the perturbations $n'$ and $T'$ from Equation \ref{03}
with the help of Equation \ref{04}. 
Then the second perturbation is absent in the resulting equation.
Finally we get
\begin{eqnarray}
{\it \Gamma}^{\,3} \,
  &-& \left[ \,
    \frac{ 1 }{ \tau_\sigma - \tau_\eta }
    \left( 1 + \left({ \frac{1}{\gamma-1} - \frac{\tau_\kappa}{\tau_\sigma} }\right)^{-1} \right) \,+\,
     \frac{ - \alpha }{ \tau_\lambda } \left({ \frac{1}{\gamma-1} - \frac{\tau_\kappa}{\tau_\sigma} }\right)^{-1} 
    \right]
    {\it \Gamma}^{\,2}   \nonumber  \\
    &+&\left[ \,\frac{ 1 }{ \tau_\sigma - \tau_\eta } \,\, \frac{ \beta - \alpha }{ \tau_\lambda } 
   \left({ \frac{1}{\gamma-1} - \frac{\tau_\kappa}{\tau_\sigma} }\right)^{-1}
   \right] {\it \Gamma} = 0 \, .
    \label{07}
\end{eqnarray}
Here the growth rate of the instability is $ {\it \Gamma} = - i \omega $.
Positive values of $ {\it \Gamma} $ correspond to the exponential growth of the perturbation in time 
while negative values indicate stabilization of the initial perturbation.
Also we introduce the new notations for the logarithmic derivatives of the cooling function
\begin{equation}
    \alpha
  = \pder{ {\rm \, ln} \, \lambda }{ {\rm \, ln} \, T } \, ,
  \qquad
  \beta
  = \pder{ {\rm \, ln} \, \lambda }{ {\rm \, ln} \, n } \, , 
    \label{08}
\end{equation}
and the characteristic times
\begin{equation}
    \tau_\sigma 
  =  \frac{ \mu \, \nu_m}{2k_{_{ \rm B }} T} \, ,
  \qquad
  \tau_\eta
  = \frac{ \eta + \nu }{ 2k_{_{ \rm B }} T n } \, ,
\qquad
\tau_\kappa
  = \frac{ \mu \, \kappa }{ (2k_{_{ \rm B }})^2 T n } \, ,
\qquad
    \tau_\lambda
  =  \frac{ 2 k_{_{ \rm B }} T n }{ \lambda } \, ,
    \label{09}
\end{equation}
of the magnetic resistivity, viscosity, thermal conduction, and optically thin radiation, respectively.
Derivatives of the heating function do not affect the development of the instability, 
as it is taken constant in time in this work.
The magnetic viscosity is denoted as
\begin{displaymath}
    \nu_m
  =  \frac{ c^2 }{4 \pi \sigma} \, .
\end{displaymath}
We mention that the Equations for characteristic times \ref{09}
are written in such a manner to provide the clarity of the final result.
For this reason, they do not coincide, 
for example, with similar equations in \citet{1982SoPh...75..237S}.
%
%

We also introduce the dimensionless parameter
\begin{equation}
\delta = 
         \left({ \frac{1}{\gamma-1} - \frac{\tau_\kappa}{\tau_\sigma} }\right)^{-1} \, ,
    \label{10}
\end{equation}
and effective viscous time
\begin{equation}
\tau_\nu =  
         { \tau_\sigma - \tau_\eta } \, .
    \label{11}
\end{equation}
Then, Equation \ref{07} can be written in the following simple form
\begin{equation}
{\it \Gamma}^{\,3} \,
  - \left[ \, 
\frac{ 1 + \delta }
         { \tau_\nu } \,+\,
     \frac{ - \alpha \, \delta }{ \tau_\lambda } \,
        \right]
    {\it \Gamma}^{\,2}
    + \left[ \, \frac{ 1 }{ \tau_\nu } \,\,
       \frac{ (\beta - \alpha) \, \delta }{ \tau_\lambda } \,\right] {\it \Gamma} = 0 \, .
    \label{12}
\end{equation}
%
%
\subsection{Features of the Instability}
\label{sec2.2}

\begin{figure}
\begin{center}
\includegraphics*[width=\linewidth]{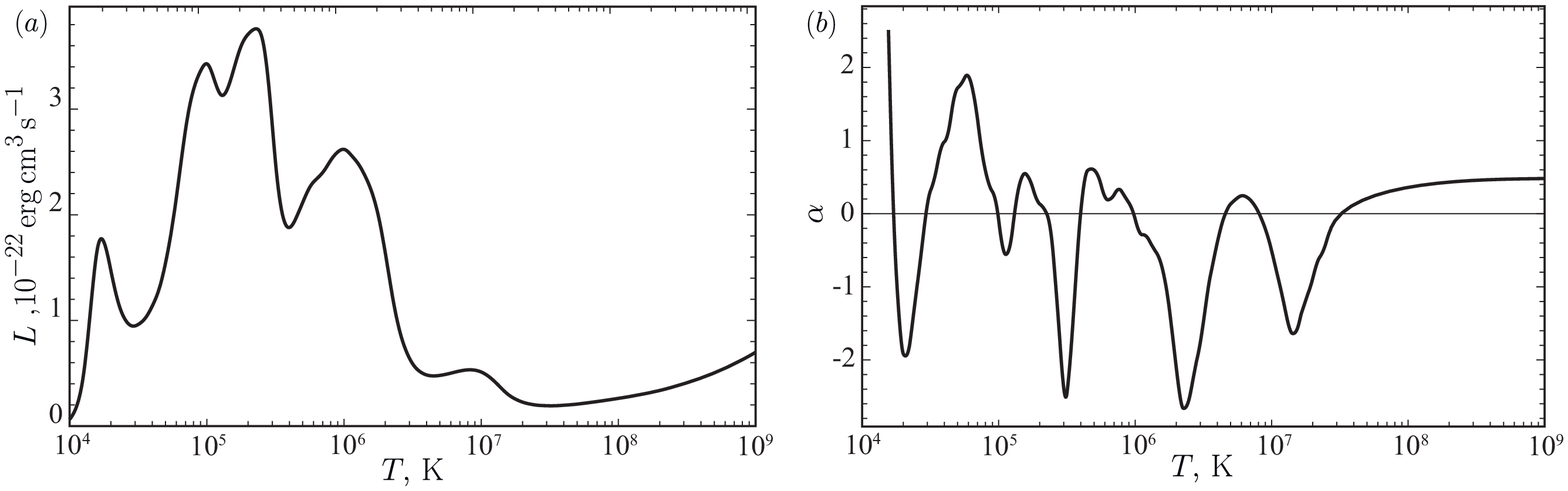}
\end{center}
\caption{(a) Radiative loss function of an optically thin medium
$L(T)$ based on the CHIANTI atomic database \citep{2019ApJS..241...22D}
for the coronal abundance of elements \citep{2012ApJ...755...33S} and $n=10^8 {\rm cm}^{-3}.$
(b) The logarithmic derivative of the radiative cooling function $\alpha$ 
with respect to the logarithm of temperature for the same conditions.}
\label{fig1}
\end{figure}
In the current section, Equation \ref{12} is applied to the physics of solar flares. 
To this end, the characteristic values of the quiet coronal plasma are used as a starting point:
$n=10^8 {\rm \, cm}^{-3}$, $T=10^6 {\rm \, K}$. 
The same instability will be considered 
in broad intervals of 
plasma densities and temperatures
in Section~\ref{sec4}.
Anomalous conductivity $\sigma=10^{12} {\rm \, s}^{-1}$ caused mainly by the ion-acoustic turbulence 
is usually applied in the context of the emerging preflare current layer \citep{2006ASSL..341.....S}.
Viscosity changes the effective viscous time according to Equation \ref{11}.
The coefficient of dynamic viscosity is estimated as \citep{1986ApJ...306..730H}
\begin{displaymath}
    \eta
    \approx 10^{-16} \, T^{\, 5/2} \, .
\end{displaymath}
Hereinafter, all quantities are measured in Gaussian units in practical equations.
Using Equations \ref{09} we are convinced that $\tau_\eta \ll \tau_\sigma$ 
here and for all further calculations in the article.
Therefore we set $\eta=0$, $\nu=0$ in what follows.
We also use a common representation of the radiative cooling function $\lambda \,(n,T)=n^2L(T)$,
where $L(T)$ is 
the radiative loss function
of an optically thin medium. 
Figure \ref{fig1}a shows the function $L(T)$ based on the 
CHIANTI version 9
atomic database \citep{2019ApJS..241...22D}
for coronal abundance elements 
(see file ${\rm sun\_coronal\_2012\_schmelz\_ext.abund}$ in the standard CHIANTI distribution and
\citeauthor{2012ApJ...755...33S}, \citeyear{2012ApJ...755...33S}).
The temperature dependence of the coefficient $\alpha$ is shown in Figure \ref{fig1}b.
The plasma thermal conductivity is considered as a free parameter in this section.

\begin{figure}
\begin{center}
\includegraphics*[width=\linewidth]{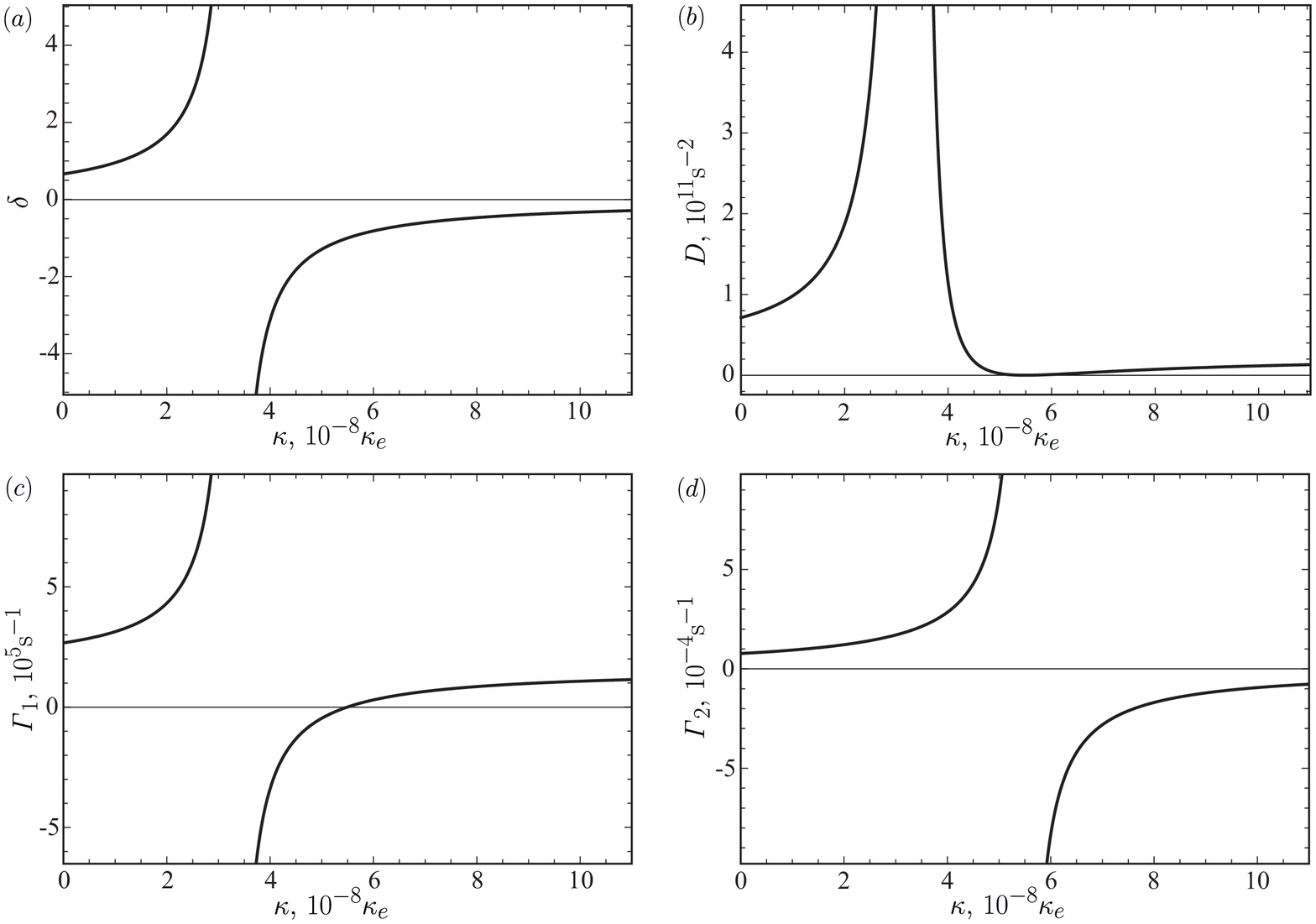}
\end{center}
\caption{
Profiles depending on the thermal conductivity of the plasma:
(a) parameter $\delta$ (Equation \ref{10}), (b) discriminant $D$,
(c) root ${\it \Gamma}_1$, and (d) root ${\it \Gamma}_2$ of Equation \ref{12}.
Thermal conductivity is measured in units of the
classical electronic thermal conductivity
calculated for $T=10^6$ K
\citep{1953PhRv...89..977S}.}
\label{fig2}
\end{figure}

The roots of Equation \ref{12} 
depend on the dimensionless parameter $\delta$.
The characteristic time $\tau_\kappa$ is directly proportional to the coefficient of thermal conductivity $\kappa$, 
while the characteristic time $\tau_\sigma$ is inversely proportional to the electrical conductivity $\sigma$
according to the definitions in Equations \ref{09}.
Therefore the fraction $\tau_\kappa / \tau_\sigma$ is proportional to
both thermal and electrical conductivities of the plasma in Equation \ref{10}.
In addition, both of them have similar physical nature 
associated with the mean free path of the particles.
It is expected that both of them increase or decrease 
under similar conditions in the plasma.
For simplicity, in this section we treat electrical conductivity as a constant
and vary thermal conductivity.
Figure \ref{fig2}a shows the dependence of the parameter $\delta$
on the thermal conductivity measured in units 
of the Spitzer's thermal conductivity \citep{1953PhRv...89..977S}
\begin{displaymath}
    \kappa_{\, \rm e}
    \approx 9 \times 10^{-7} \, T^{5/2} \, .
\end{displaymath}

As one can see, $|\,\delta\,|<1$ for all $\kappa$, 
except an interval $2 \times 10^{-8} \,\kappa_{\, \rm e} < \kappa < 6 \times 10^{-8} \,\kappa_{\, \rm e}$.
The sign of the parameter $\delta$ changes 
when the plasma thermal conductivity decreases
to $\kappa \lesssim 3 \times 10^{-8} \,\kappa_{\, \rm e}$.
For example, if thermal conductivity is suppressed 
by a perturbation of the magnetic field, 
then ionic thermal conduction becomes more efficient \citep{1958PhRv..109....1R}
\begin{displaymath}
    \kappa_{\, \rm i}
    \approx 2 \times 10^{-17} \,
    \frac{ n^2 }{ T^{1/2} B'^2 } \, .
\end{displaymath}
So in Section \ref{sec3},
the thermal conductivity inside the current layer is suppressed 
by a perturbation of the magnetic field
directed along the external magnetic field
(for more details on the field configuration see Section \ref{sec3}).
The magnitude of the required perturbation 
of the magnetic field can be found from the evaluation
$\kappa_{\, \rm i} \approx 3 \times 10^{-8} \,\kappa_{\, \rm e}$.
The amplitude of the magnetic field perturbation $B'~\gtrsim~0.01~{\rm \, G}$
is sufficient to change the sign of the parameter $\delta$.
This will be important for further discussion.
%
%

The roots of Equation \ref{12} are as follows:
\begin{displaymath}
    {\it \Gamma}_0 = 0 \, ,
\end{displaymath}
\begin{displaymath}
    {\it \Gamma}_{1,2} = \frac{1}{2}\left\{
    \left( \, 
\frac{ 1 + \delta }
         { \tau_\nu } \,+\,
     \frac{ - \alpha \, \delta }{ \tau_\lambda } \,
        \right) \pm \left[ \left( \, 
\frac{ 1 + \delta }
         { \tau_\nu } \,+\,
     \frac{ - \alpha \, \delta }{ \tau_\lambda } \,
        \right)^2 - 4 \, \frac{ 1 }{ \tau_\nu } \, \frac{ (\beta - \alpha) \, \delta }{ \tau_\lambda }  \right]^{1/2}\right\}
\end{displaymath}
The root ${\it \Gamma}_0$ is not of interest here, 
since it corresponds to the transition to a new stationary state, 
which differs from the initial one by the magnitude of the perturbation.
The relation $\tau_\nu / \tau_\lambda$ is much smaller than 1 for the described conditions of the solar corona.
Therefore, the roots ${\it \Gamma}_{1,2}$ can be expanded in small parameter $\tau_\nu / \tau_\lambda $.
Keeping only zero-order terms, one obtains:
\begin{equation}
    {\it \Gamma}_1 \simeq
    \frac{ 1+\delta }{ \tau_\nu } \, ,
    \qquad
        {\it \Gamma}_2 \simeq
    \frac{ \beta-\alpha }{ \tau_\lambda } \, \frac{ \delta }{ 1+\delta } \, .
    \label{13}
\end{equation}
%
%

Figure \ref{fig2}b shows the dependence of the discriminant $D$ 
in Equation \ref{12} on the thermal conductivity, while
Figure \ref{fig2} c and d shows the roots ${\it \Gamma}_{1,2}$.
The exact calculation of the roots ${\it \Gamma}_{1,2}$ completely coincides 
with the approximate Equations \ref{13} in the scale of Figure \ref{fig2} c and d.
Differences are observed only
in the region of rapid growth of $|\,\delta\,|$, 
where the discriminant $D$ also tends to infinity,
and in the region where the discriminant $D$ is negative (Figure \ref{fig2}b).
The root ${\it \Gamma}_{1}$ has a discontinuity in the first region (Figure \ref{fig2}c), 
while the root ${\it \Gamma}_{2}$ has a discontinuity in the second region (Figure \ref{fig2}d).
In these areas, the linear approximation of the problem of small perturbations is unsuitable.
Changing the initial parameters $n$, $T$, and $\sigma$ 
within the limits which are acceptable for the conditions of the solar corona 
stretches or compresses Figure \ref{fig2} along the coordinate axes, 
but does not make any qualitative changes in these plots.
%
%

The figure shows that ${\it \Gamma}_1 \gg {\it \Gamma}_2$ for almost all values $\kappa$
except for a narrow interval near $\kappa = 4 \times 10^{-8} \,\kappa_{\, \rm e}$
where ${\it \Gamma}_1$ is negative and ${\it \Gamma}_2$ is positive. 
This means that the instability described by the root ${\it \Gamma}_1$ should grow
much faster than the instability described by the root ${\it \Gamma}_2$
everywhere except in this narrow interval. 
In the geometry of the preflare current layer in the solar corona, 
the spatial scale of the root ${\it \Gamma}_1$ does not satisfy the MHD approximation used (Section \ref{sec4.1})
and we should use a higher frequency approximation to study it further.
Therefore, in what follows, we will focus on the root ${\it \Gamma}_2$ and assume that 
the value of the thermal conductivity satisfies the condition ${\it \Gamma}_2 > {\it \Gamma}_1$.

\section{Current Layer Model}
\label{sec3}

\begin{figure}
\begin{center}
\includegraphics*[width=0.75\linewidth]{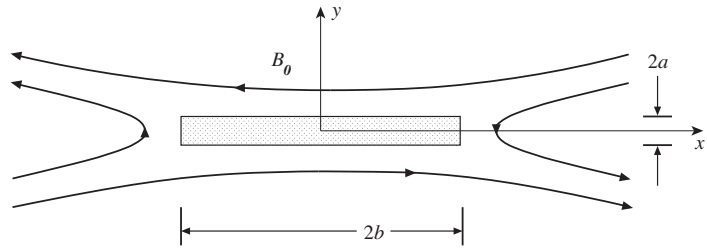}
\end{center}
\caption{Location of the current layer in the coordinate system.}
\label{fig3}
\end{figure}

We consider the piecewise homogeneous model 
of the preflare current layer, presented by \cite{1982SoPh...75..237S}.
The current layer is located in the $(x,z)$ plane (Figure \ref{fig3}).
The $z$-axis complements the right triplet $(x,y,z)$ and is directed toward the reader in Figure \ref{fig3}. 
The plasma concentration and temperature inside the layer 
are equal to $n_s$ and $T_s$, respectively.
The current layer is assumed magnetically neutral, $B_s = 0$, 
without any directed plasma flows, i.e. $v_s = 0$.
The half-thickness of the current layer $a$ 
is much smaller than its half-width $b$.
When considering the preflare non-reconnecting current layer, 
$b \to \infty$ is assumed.
As a consequence, $\partial / \partial {x}=0$ in such model.
This means that
we neglect the evolution of the current layer
along the $x$-axis, such as the tearing instability \citep{1963PhFl....6..459F}.
We focus on the structure of the current layer along the $z$-axis.
The inner region of the current layer is separated from the outer plasma 
by a tangential discontinuity \citep{2015AdSpR..56.2779L}.
Outside the layer, we denote the concentration and the temperature of the homogeneous plasma
as $n_0$ and $T_0$, correspondingly.
A uniform magnetic field $B_0$ is directed against the $x$-axis for positive $y$ 
and along the $x$-axis for negative $y$.
Thus, the current in the layer is directed along the $z$-axis.
In order to study the effect of thermal balance on the structural stability 
of a preflare current layer, the effects of viscosity, electrical and thermal conductivity, 
and radiative cooling are considered inside the current layer, 
but these effects are insignificant outside.
An important difference between the model considered here 
and the \cite{1982SoPh...75..237S} model is the possibility 
of penetration of a magnetic field perturbation inside the current layer.
Mathematically, this comes to considering the current layer interior
in the magnetohydrodynamic approximation rather than in the hydrodynamic one.
%


\subsection{Outside the Current Layer}
\label{sec3.1}

Following \cite{1982SoPh...75..237S}, 
we set $\sigma \to \infty$, $\kappa=0$, $\lambda=0$, $\eta=0$, and $\nu=0$ 
in the set of Equations \ref{01} outside the current layer.
Plasma density contrast inside and outside the super-hot turbulent-current layers is about 5 
(see Section 8.5.3 in \citeauthor{2006ASSL..341.....S}, \citeyear{2006ASSL..341.....S}).
Kinetic models give the same values \citep{2015PhPl...22k2902K, 2017A&A...600A..78P}.
Wherein, the external plasma could radiate up to a factor of 100 less efficiently than the internal one. 
In other words, the characteristic timescales of radiative processes outside the layer and those inside it 
(including the characteristic timescales of the perturbation and of the other non-adiabatic processes) 
could differ by two orders of magnitude, allowing one to neglect the effects of radiation in the external plasma.
Moreover, we suppose that the considered preflare current layer is more similar 
to the neutral current layer by Syrovatskii, 
in which the density contrast can be much higher \citep{1976SvAL....2...13S}.
In addition, the plasma is assumed to be at rest, i.e. $v_0=0$.
The solution is sought in the form of a periodic perturbation
along the $z$-axis in Figure \ref{fig3} which decays exponentially with distance from the current layer:
\begin{displaymath}
    f(y,z,t)
  = f_0
  + f_1(y) \, {\rm exp} \, ( - i \omega t + i k_z z ) \, ,
\end{displaymath}
\begin{displaymath}
    f_1(y) _{\rm top}=f_{\rm 1 \, top} \, {\rm exp} \, [ - k_{y1} (y-a) ] \, , 
    \qquad
    f_1(y) _{\rm bottom}=f_{\rm 1 \, bottom} \, {\rm exp} \, [ k_{y1} (y+a) ] \, ,
\end{displaymath}
with perturbation amplitudes
\begin{displaymath}
    f_{\rm 1 \, top} \equiv \{{v_{y1}, v_{z1}, n_1, T_1, B_{x1}}\} \, ,
    \qquad
    f_{\rm 1 \, bottom} \equiv \{{-v_{y1}, v_{z1}, n_1, T_1, -B_{x1}}\} \, ,
\end{displaymath}
on either side outside the current layer, respectively.
Here, $\omega$ is the perturbation frequency, 
$k_z$ and $k_{y1}$ are the perturbation wave numbers along the $z$ and $y$ axes, respectively, 
and $a$ is the half-thickness of the current layer.
Index ``$1$'' refers to quantities outside the layer.
Thus, we are looking for a solution in the form of a perturbation 
that propagates 
through the surface of the current layer 
and decays with distance from it.

Based on the symmetry of the problem, 
the set of Equations \ref{01} is considered only for the upper half space.
Neglecting the squares of the perturbed quantities, 
one finds the linearized system of equations:
\begin{equation}
    i \omega \, n_1
  = - k_{y1} \, n_0 v_{y1} + i k_z \, n_0 v_{z1} \, ,
    \label{14}
\end{equation}
\begin{equation}
    i \omega \, \mu n_0 v_{y1}
  = - k_{y1} \, 2k_{_{ \rm B }}  (n_0 T_1 + T_0 n_1)
    + k_{y1} \, \frac{ B_0  }{ 4 \pi }\,B_{x1}  \, ,
    \label{15}
\end{equation}
\begin{equation}
    i \omega \, \mu n_0 v_{z1}
  = i k_{z} \, 2k_{_{ \rm B }}  (n_0 T_1 + T_0 n_1)
  - i k_{z} \, \frac{ B_0  }{ 4 \pi } \, B_{x1} \, ,
    \label{16}
\end{equation}
\begin{equation}
    ( \gamma - 1 ) \, T_0 n_1 = n_0 T_1 \, ,
    \label{17}
\end{equation}
\begin{equation}
    i \omega \, B_{x1}
  = k_{y1} \, B_0 v_{y1} - i k_z \, B_0 \, v_{z1} \, .
    \label{18}
\end{equation}
The dispersion relation for perturbations outside the current layer is determined by
equating the determinants of a homogeneous system of linear Equations \ref{14}\,--\,\ref{18} to zero 
\begin{equation}
    k_{y1}^2 = k_z^2 - \frac{\omega^2}{ V_S^2 + V_A^2 } \, ,
    \label{19}
\end{equation}
where the sound and the Alfv\'en speeds are denoted as
\begin{equation}
    V_S = \sqrt{ \frac{ 2 \gamma k_{_{ \rm B }} T_0 }{ \mu }} \, ,
    \qquad
        V_A = \frac{B_0}{ \sqrt{ 4 \pi n_0 \mu } } \, ,
    \label{20}
\end{equation}
respectively.
The dispersion relation in Equation \ref{19} describes a fast magnetoacoustic wave 
propagating over the surface of the current layer perpendicular to the magnetic field.
%
%

\vspace{1mm}

%
%
\subsection{Inside the Current Layer}
\label{sec3.2}

Dissipative effects of Joule and viscous heating, 
thermal conductivity, and radiative cooling 
should be considered inside the current layer.
The current layer is assumed magnetically neutral, $B_s=0$, 
without any directed plasma flows, i.e. $v_s=0$.
The solution is sought in the same form as outside the current layer
\begin{displaymath}
    f(y,z,t) = f_s
             + f_2(y) \, {\rm exp} \, (-i\omega t + ik_zz) \, .
\end{displaymath}
The perturbations decrease exponentially along the $y$-axis 
when moving from the upper boundary of the current layer 
\begin{displaymath}
    f_2(y)_{\rm top} =f_{\rm 2 \, top} \, {\rm exp} \, [ - k_{y2} (a - y) ] \, ,
\end{displaymath}
\begin{displaymath}
    f_{\rm 2 \, top} \equiv \frac{1}{2} \, {\rm exp} (k_{y2} a) \, \{ v_{y2}, v_{z2}, n_2, T_2, B_{x2} \} \, 
\end{displaymath}
and when moving from the lower boundary
\begin{displaymath}
    f_2(y)_{\rm bottom} =f_{\rm 2 \, bottom} \, {\rm exp} \, [ - k_{y2} (a + y) ] \, ,
\end{displaymath}
\begin{displaymath}
    f_{\rm 2 \, bottom} \equiv \frac{1}{2} \, {\rm exp} (k_{y2} a) \, \{ -v_{y2}, v_{z2}, n_2, T_2, -B_{x2} \} \, .
\end{displaymath}
Here, index ``$2$'' refers to perturbations inside the layer, 
$\frac{1}{2} \, {\rm exp} (k_{y2} a)$ is a scale factor that unifies the solution for all thicknesses of the current layer.
Inside the current layer, perturbations coming from the upper and lower boundaries add up
\begin{displaymath}
f_2(y) = f_2(y)_{\rm top} + f_2(y)_{\rm bottom} \, .
\end{displaymath}

The resulting dependences of the perturbations on the coordinate $ y $ 
are hyperbolic functions.
The sum of the perturbations which are odd in the $y$ direction gives a hyperbolic sine
\begin{displaymath}
\left\{
\begin{array}{c}
v_{y2}(y)\\
B_{x2}(y)
\end{array}
\right\} =
\left\{
\begin{array}{c}
v_{y2}\\
B_{x2}
\end{array}
\right\}
{\rm sinh} \, (k_{y2} y) \, ,
\end{displaymath}
The sum of the perturbations which are even in $y$ gives a hyperbolic cosine
\begin{displaymath}
\left\{
\begin{array}{c}
v_{z2}(y)\\
n_2(y)\\
T_2(y)
\end{array}
\right\} =
\left\{
\begin{array}{c}
v_{z2}\\
n_2\\
T_2
\end{array}
\right\}
{\rm cosh} \, (k_{y2} y) \, .
\end{displaymath}
Real values of $k_{y2}$ determine the effective thickness of the skin depth of the layer, 
that is, the distance to which the disturbance of the layer boundary penetrates.
On the other hand, a standing wave is formed inside the current layer along the $y$-axis at imaginary $k_{y2}$.
The value of the wave number $k_{y2}$ is not prescribed and can be determined from the solution, 
however, this is not the purpose of this article.
The solution describes the plasma motion which is symmetric about the $(x,z)$ plane.
%
%

The set of Equations \ref{01} is linearized as in the previous section:
\begin{equation}
    i \omega \, n_{2}
  = k_{y2} \, n_s v_{y2} + i k_z \, n_s v_{z2} \, ,
    \label{21}
\end{equation}
\begin{equation}
    i \omega \, \mu n_s v_{y2}
  = k_{y2} \, 2 k_{_{ \rm B }} ( n_s T_2 + T_s n_2 ) + (k_z^2-k_{y2}^2) \, 
  \eta v_{y2} - i \omega k_{y2} \frac{\nu}{n_s} n_2 \, ,
    \label{22}
\end{equation}
\begin{equation}
    i \omega \, \mu n_s v_{z2}
  = i k_{z} \, 2 k_{_{ \rm B }} ( n_s T_2 + T_s n_2 ) + (k_z^2-k_{y2}^2) \, 
  \eta v_{z2} - i \omega i k_{z} \frac{\nu}{ n_s} n_2 \, ,
    \label{23}
\end{equation}
\begin{equation}
    i \omega \, \frac{ 2 k_{_{ \rm B }} n_s }{ \gamma - 1} \, T_2
  - i \omega \, 2 k_{_{ \rm B }} T_s \, n_2 =  ( k_z^2 - k_{y2}^2 ) \, \kappa T_2
   + \pder{ \lambda }{ T } \, T_2 + \pder{ \lambda }{ n } \, n_2 \, ,
    \label{24}
\end{equation}
\begin{equation}
    i \omega \, B_{x2}
  = (k_z^2 - k_{y2}^2) \, \nu_m B_{x2} \, .
    \label{25}
\end{equation}
Equation \ref{25}, like the set of Equations \ref{05} and \ref{06}, can be satisfied at $B_{x2}=0$, 
however, if the perturbation of the magnetic field, $B_{x2}\ne0$, penetrates into the current layer, 
then Equation \ref{25} gives an additional dispersion relation independent of the $B_{x2}$ value.
After expressing the difference $k_z^2 - k_{y2}^2$ from Equation \ref{25}, 
we can exclude it from the remaining Equation \ref{21}\,--\,\ref{25}.
Making transformations similar to those in Section \ref{sec2.1}, 
gives again Equation \ref{12} with $n=n_s$ and $T=T_s$.
It is worth noting that the possibility to directly determine the instability increment 
is due to the simultaneous presence of two dispersion relations in Equations \ref{21}\,--\,\ref{25} at once: 
the first follows from Equations \ref{21}\,--\,\ref{24}, and the second one from Equation \ref{25}.
An important feature of Equations \ref{21}\,--\,\ref{25}
is that the wave vector $k_z$ turns out to be perpendicular 
to the arising perturbation of the magnetic field $B_{x2}$.
This promotes the suppression of thermal conductivity 
along the $z$-axis (in the direction of the current)
and the formation of a thermal instability (see Section \ref{sec2.2}).
%
%

\vspace{1mm}

%
%
\subsection{Boundary of the Current Layer}
\label{sec3.3}

The considered model of the current layer has no 
plasma motion neither outside nor inside the layer
in equilibrium
($v_0=0, v_s=0$), 
but a magnetic field jump occurs at the layer boundary.
A tangential discontinuity in MHD corresponds to such conditions \citep{2015PhyU...58..107L}.

The sum of the gas-dynamic and magnetic pressure should be equal 
on the different sides of the tangential discontinuity \citep{1956TrFIAN...13..64S}.
In a linearized form, it looks as follows
\begin{equation}
    n_0 T_1 + T_0 n_1 - \frac{ B_0 B_{x1} }{ 8 \pi k_{_{ \rm B }} }
  = ( n_s T_2 + T_s n_2 ) \, {\rm cosh} \, ( k_{y2} a ) \, .
    \label{26}
\end{equation}
The left side of Equation \ref{26} can be expressed in terms of the perturbation $v_{y1}$
using Equation \ref{15}.
The right side of Equation \ref{26} can be expressed in terms of the perturbation $v_{y2}$
using Equations \ref{21}\,--\,\ref{23}
and substituting $k_z^2 - k_{y2}^2$ from Equation \ref{25}.
Then Equation \ref{26} takes the form
\begin{equation}
    - \, \frac{n_0}{n_s} \, \frac{v_{y1}}{k_{y1}} 
  = \frac{\tau_\nu}{\tau_\sigma} \, \frac{v_{y2}}{k_{y2}} \, {\rm cosh} \, ( k_{y2} a ) \, .
    \label{27}
\end{equation}
Here, the Equations \ref{09} are also used with substitutions $n=n_s$ and $T=T_s$.

Velocity perturbations distort the surface of the tangential discontinuity. 
For reasons of continuity, the velocity perturbation on both sides of the discontinuity 
should have the same magnitude and direction
\begin{equation}
    v_{y1}^2 + v_{z1}^2 = v_{y2}^2 \, {\rm sinh}^2 \, ( k_{y2} a ) + v_{z2}^2 \, {\rm cosh}^2 \, ( k_{y2} a ) \, ,
    \label{28}
\end{equation}
\begin{equation}
    \frac{v_{z1}}{v_{y1}} = \frac{v_{z2} \, {\rm cosh} \, ( k_{y2} a )}{v_{y2} \, {\rm sinh} \, ( k_{y2} a )} \, .
    \label{29}
\end{equation}
Equation \ref{28} is then rewritten as
\begin{equation}
    v_{y1} = \pm \, v_{y2} \, {\rm sinh} \, ( k_{y2} a ) \, ,
    \label{30}
\end{equation}
where the choice of sign depends on the signs of perturbations $v_{y1}$ and $v_{z1}$.

Let us divide Equation \ref{30} by Equation \ref{27}
\begin{equation}
    \pm \, \frac{\tau_\nu}{\tau_\sigma} \, \frac{n_s}{n_0} \, k_{y1} = k_{y2} \, {\rm tanh} \, ( k_{y2} a ) \, .
    \label{31}
\end{equation}
Equation \ref{31} differs from Equation 23 of \cite{1982SoPh...75..237S} 
by a coefficient $ \tau_\nu / \tau_\sigma$ determining the role of viscosity
in the formation of the structure of the preflare current layer.
Note that the right side of Equation \ref{31} is positive for any real $k_{y2}$.
This means that $ \pm(\tau_\nu / \tau_\sigma)$ should also be positive 
for physically meaningful values $k_{y1}$.
Substitution of the wave numbers $k_{y1}$ and $k_{y2}$ 
from Equations \ref{19} and \ref{25}, respectively, 
gives the dispersion relation 
which relates the instability increment to the wave number $k_z$
\begin{equation}
    \left( \frac{\tau_\nu}{\tau_\sigma} \,
    \frac{n_s}{n_0}\right)^2
    \left[ k_z^2 + \frac{ {\it \Gamma}^2 }{ V_S^2 + V_A^2 } \right]
  = \left[ \, k_z^2 +\frac{ {\it \Gamma} }{ \nu_m } \,
    \right]
    {\rm tanh}^2
    \left\{ a
    \left[ \, k_z^2 + \frac{ {\it \Gamma} }{ \nu_m } \,
    \right]^{1/2}
    \right\} \, .
    \label{32}
\end{equation}
We determine the growth increment, $ {\it \Gamma} = - i \omega $, from Equation \ref{12}, 
which is identical to the solution of Equations \ref{21}\,--\,\ref{25}.
Then we determine the spatial period of the
instability, $ l = 2 \pi / k_z $, from Equation \ref{32}.

\section{Thermal Instability of the Current Layer}
\label{sec4}

Now the growth rate of the instability can be calculated using Equation \ref{12} 
and the corresponding spatial period of the perturbation can be found from Equation \ref{32}. 
For this aim, the appropriate values of the parameters 
of the current layer and the surrounding plasma should be chosen.
Taking into account the possibility of plasma gathering by magnetic fields 
and its heating during the formation of the current layer
before the onset of the studied instability, 
the range of values $n_0=10^8-10^{12} {\rm \, cm}^{-3}$,
$n_s / n_0=10-10^3$, $T_0=10^6 {\rm \, K}$, $T_s=10^6-10^8 {\rm \, K}$, 
$B_0=1-10^2 {\rm \, G}$, $\sigma=10^{11} {\rm \, s}^{-1}$, $a=10^5-10^7 {\rm \, cm}$
is considered.
This range covers all the reasonable parameters of the coronal plasma.
As one can see from the first two brackets on the left hand side of Equation \ref{32}, 
the effect of an increase of viscosity is the opposite to an increase in the density jump. 
Therefore, no viscosity is introduced, because
its effect is taken into account in the density jump ($\eta=0$, $\nu=0$).
In addition, the viscosity effect is small ($\tau_\nu \approx \tau_\sigma$)
in the investigated range of coronal plasma parameters.

\subsection{Growth Increment in the Current Layer}
\label{sec4.1}

The instability occurs when the roots of Equation \ref{12} are positive.
The roots ${\it \Gamma}_1$ and ${\it \Gamma}_2$ are real numbers 
everywhere except for a narrow interval where $D<0$ (Figure \ref{fig2}b). 
In this interval, the roots become complex.
Three unstable solutions are possible: 
the left branch of ${\it \Gamma}_1$ (Figure \ref{fig2}c), 
the positive part of the right branch of ${\it \Gamma}_1$ (also Figure \ref{fig2}c), 
and the left branch of ${\it \Gamma}_2$ (Figure \ref{fig2}d).
As one can see, $|\,{\it \Gamma}_1\,| \gg |\,{\it \Gamma}_2\,|$ everywhere 
except perhaps in a small area near $D=0$ (see Figure \ref{fig2}b).
Calculation of the instability scale over the entire range 
of coronal plasma parameters described above 
gives $l\lesssim10^4 {\rm \, cm}$.
It is less than the corresponding Larmor radius of the proton
for most of the values of plasma parameters.
Moreover, complex values of ${\it \Gamma}_1$ for $D<0$ lead to complex values of $k_z$, 
which corresponds to the spatial attenuation of the perturbation at the same scales 
($l\lesssim10^4 {\rm \, cm}$).
The presence of viscosity can only increase the value ${\it \Gamma}_1$ 
as seen from Equations \ref{11} and \ref{13}. 
Therefore, it further reduces the scale of instability.

The MHD approximation is incorrect for the description of the plasma at such scales.
Therefore, in this article we cannot say whether such instability appears in a more general kinetic description. 
Remaining within the framework of the MHD, we further consider the root ${\it \Gamma}_1$ physically meaningless.
In any case, if the instability associated with the root ${\it \Gamma}_1$ exists 
in the kinetic description and dominates the instability with an increment ${\it \Gamma}_2$, 
there is a narrow interval of plasma thermal conductivities $\kappa$ 
where ${\it \Gamma}_1$ is negative and ${\it \Gamma}_2$ is positive,
and an instability occurs due to the root ${\it \Gamma}_2$ (Figure \ref{fig2}).
We assume that the thermal conductivity is suppressed 
by the perturbation of the magnetic field in the current layer, 
which triggers the instability. 
Note that, for further reasoning, it is not important 
which process led to the suppression of the thermal conductivity. 
The space scale of the instability \ref{32} does not depend 
on the exact value of the thermal conductivity coefficient
and can be calculated for any range of coronal plasma parameters.

The negative right branch of the root ${\it \Gamma}_2$ indicates 
the stabilizing effect of the high thermal conductivity of the plasma.
However, if, for some reason,
the thermal conductivity falls below the threshold value 
$\delta=-1$ (see Equation \ref{13} and Figure \ref{fig2}d), 
${\it \Gamma}_2$ becomes positive and an instability occurs.
The value $\beta-\alpha$ is positive over the entire range 
of the above-described coronal plasma conditions.
As it was shown in Section \ref{sec2.2}, the transverse magnetic field 
can cause a decrease of the thermal conductivity.
Equations \ref{21}\,--\,\ref{25} allow the
perturbations of the $x$-component of the magnetic field 
to appear inside the current layer. 
This field is actually perpendicular to $\nabla T$
in the current layer under consideration.
However, the specific nature of the suppression of the thermal conductivity 
is not very important for further considerations.
It is enough for us to assume that the thermal conductivity 
went down for some reason ($\tau_\kappa\ll\tau_\sigma$).
Then the growth rate of the instability tends to the value
\begin{equation}
   {\it \Gamma} =
    \frac{ 2 }{ 5 } \, \frac{ \beta-\alpha }{ \tau_\lambda } 
    \label{33}
\end{equation}
(see Equations \ref{13} and \ref{10}).
The growth time of the instability is proportional to the characteristic time of the
plasma cooling and depends 
on the logarithmic derivatives of the radiative cooling function 
(with respect to concentration and temperature).
Thermal instability criteria \citep{1965ApJ...142..531F} in our notation can be written as follows:
$$\alpha<0 \qquad \eqno({\rm isochoric}) \, ,$$
$$\alpha<\beta-1 \qquad \eqno({\rm isobaric}) \, ,$$
$$\alpha<-\frac{\beta-1}{\gamma-1} \qquad \eqno({\rm isentropic}) \, ,$$
where $\beta=2$, $\gamma=5/3$.
Thus, the criteria for isochoric, isobaric, and isentropic instabilities are
$\alpha<0$, $\alpha<1$, and $\alpha<-3/2$, respectively.
Figure \ref{fig1}b shows that the isobaric criterion of thermal instability is fulfilled
for the entire range of coronal plasma parameters.
We expect that the instability discussed in this work is a special case of
the condensation mode of the isobaric thermal instability.

\subsection{Spatial Period in the Current Layer}
\label{sec4.2}

\begin{figure}
\begin{center}
\includegraphics*[width=\linewidth]{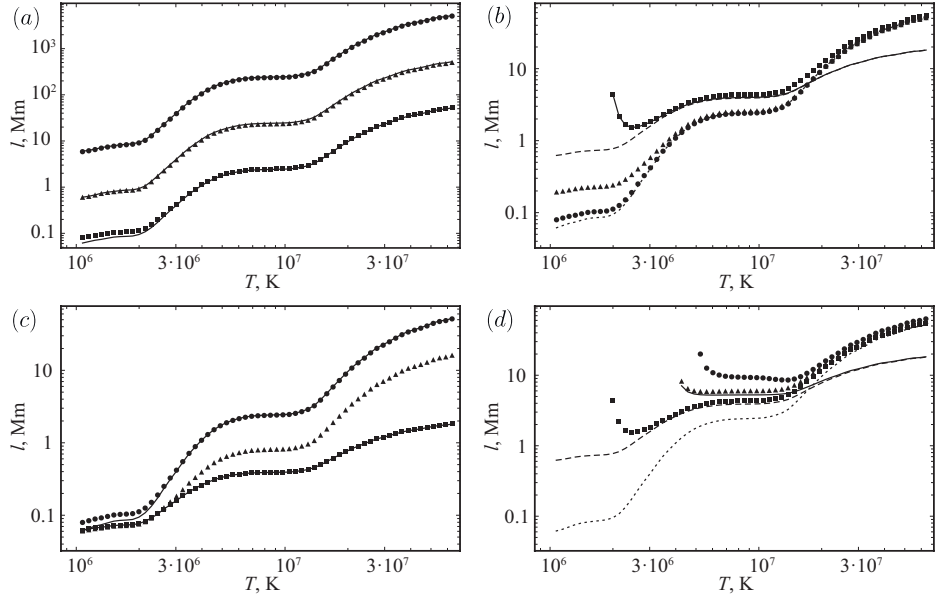}
\end{center}
\caption{The spatial period of the instability depending on the temperature of the current layer.
Parameters of the coronal plasma: $n_0=10^{10} {\rm \, cm}^{-3}$,
$n_s/n_0=10$, 
$a=10^5 {\rm \, cm}$, 
$B_0=100 {\rm \, G}$.
One of the parameters changes in each figure:
(a) $n_0=10^8 {\rm \, cm}^{-3}$ (circles), $n_0=10^9 {\rm \, cm}^{-3}$ (triangles), 
$n_0=10^{10} {\rm \, cm}^{-3}$ (squares), 
solid lines show analytical solutions (Equation \ref{35});
(b) $n_s/n_0=10$ (circles), $n_s/n_0=100$ (triangles), $n_s/n_0=1000$ (squares), 
dotted line shows 
the analytical solution
(Equation \ref{35}) for circles, solid and dashed lines show analytical solutions for 
Equation \ref{36} and \ref{37}, respectively, for squares;
(c) $a=10^5 {\rm \, cm}$ (circles), $a=3\times10^5 {\rm \, cm}$ (triangles), $a=10^7 {\rm \, cm}$ (squares),
upper and lower solid lines show analytical solutions for Equation \ref{35} and \ref{37}, respectively;
(d) $n_s/n_0=1000$, 
$B_0=1 {\rm \, G}$ (circles), $B_0=10 {\rm \, G}$ (triangles), $B_0=100 {\rm \, G}$ (squares)
dotted line shows analytical solution (Equation \ref{35}) for circles, 
solid line shows analytical solution (Equation \ref{36}) for triangles,
dashed line shows analytical solution (Equation \ref{37}) for squares.
}
\label{fig4}
\end{figure}
Equation \ref{32} has two obvious approximations: 
${\rm tanh} \, ( k_{y2} a ) \to k_{y2} a$ for small $k_{y2} a$ 
and ${\rm tanh} \, ( k_{y2} a ) \to 1$ for large $k_{y2} a$.
In what follows, they are called the thin and thick approximations, respectively. 
In the first case, the current layer is thin enough, so that the perturbation 
arising at one boundary of the layer does not decay 
along the way to the other boundary. 
On the contrary, the current layer is quite thick 
compared to the attenuation length of the perturbation in the second case.
Numerical calculations of Equation \ref{32} with ${\it \Gamma}$ from Equation \ref{33} show that
$k_z^2\ll {\it \Gamma}/\nu_m$ over the entire range of coronal plasma parameters.
Therefore, the dispersion Equation \ref{32} can be simplified.

In the thin approximation,
\begin{equation}
    k_{z{\rm thin}}^2
    \simeq \left[\left( \frac{\tau_\sigma}{\tau_\nu} \,
    \frac{n_0}{n_s}\right)^2 \frac{ 1 }{V_D^2} - 
    \frac{ 1 }{ V_S^2 + V_A^2 }\right] \, {\it \Gamma}^2 \, ,
    \label{34}
\end{equation}
where the drift velocity $V_D=\nu_m / a$ is introduced.
This is the velocity at which the plasma drifts into the current layer 
(see Section 8.1.1 in \citeauthor{2006ASSL..341.....S}, \citeyear{2006ASSL..341.....S}). 
There is no drift in our model, but we will use this notation for convenience.
For sufficiently strong magnetic field (see Equation \ref{20}) and low viscosity, 
Equation \ref{34} transforms to
\begin{equation}
    k_{z{\rm thin}}
    \simeq 
    \frac{n_0}{n_s} \, \frac{ {\it \Gamma} }{V_D} \, .
    \label{35}
\end{equation}

In the thick approximation,
\begin{equation}
    k_{z{\rm thick}}^2
    \simeq \left( \frac{\tau_\sigma}{\tau_\nu} \,
    \frac{n_0}{n_s}\right)^2 \frac{ {\it \Gamma} }{\nu_m} - 
    \frac{ {\it \Gamma}^2 }{ V_S^2 + V_A^2 } \, ,
    \label{36}
\end{equation}
and for strong field and low viscosity,
\begin{equation}
    k_{z{\rm thick}}
    \simeq 
    \frac{n_0}{n_s} \, \sqrt{\frac{ {\it \Gamma} }{\nu_m}} \, .
    \label{37}
\end{equation}

Figure \ref{fig4} shows a series of profiles of the dependence 
for the spatial period of the instability 
calculated as $l=2\pi/k_{z}$ for 
$k_z$ from Equation \ref{32} (circles, triangles, and squares)
and $k_{z{\rm thin}}$ and $k_{z{\rm thick}}$ approximations 
from Equations \ref{35} and \ref{37}, respectively (thin lines),
on the temperature of the current layer.
The exact calculation of Equations \ref{12} and \ref{32}
is shown by circles, triangles, and squares in the figure. 
The approximative Equations \ref{35}\,--\,\ref{37} are shown by solid, dashed, and dotted lines.

The spatial period of the instability 
strongly depends on the concentration of the surrounding plasma (Figure \ref{fig4}a).
The graphs are in good agreement with the thin approximation (Equation \ref{35}).
Using Equation \ref{34} instead of Equation \ref{35} does not lead to a visible improvement in the result.
The most remarkable feature of the graphs is a step at $T_s=5\times10^6-10^7 {\rm \, K}$.
The spatial period is constant in a fairly wide temperature range, 
and it is this temperature range that seems quite reasonable for a preflare current layer.
It is also reasonable to expect an increase in plasma concentration near the current layer. 
With an increase in the strength of the magnetic field 
from 1 G (for a quiet corona) to 100 G (for the active region), 
the plasma concentration also increases by two orders of magnitude due to magnetic freezing. 
Therefore, $n_0=10^{10} {\rm \, cm}^{-3}$ is used for the other graphs in Figure \ref{fig4}.

An increase in the concentration jump does not change the spatial period quantitatively, 
but it changes the solution qualitatively (Figure \ref{fig4}b).
Large density jumps at low temperatures $T_s<10^7$ correspond to a thick approximation, 
and Equation \ref{36} follows the exact solution much better than Equation \ref{37}.

An increase in the half-thickness of the current layer obviously leads to a thick approximation, 
but it also slightly changes the spatial period of instability (Figure \ref{fig4}c).

The influence of the magnetic field is manifested only at high jumps in concentration
when the second term on the right side of the Equations \ref{34} and \ref{36} prevails.
Therefore, Figure \ref{fig4}d is calculated for $n_s/n_0=1000$.
Again, the dependence of the spatial period on the magnitude of the magnetic field is rather weak.
The influence of the magnetic field becomes indistinguishable at lower density contrasts.

As a result, the spatial period of the instability is constant 
over a wide range of changes in the parameters 
of the coronal plasma at the assumed temperature 
of the preflare current layer $T_s=5\times10^6-10^7 {\rm \, K}$
and the concentration of the surrounding plasma $n_0=10^{10} {\rm \, cm}^{-3}$. 
Its values belong to in a narrow range from 1 to 10 Mm, 
which is in good agreement with the distances between the solar flare loops
observed in the ultraviolet range.

\section{Conclusion}
\label{sec5}

The stability problem of the preflare current layer with respect to small perturbations is addressed.
The problem is solved within the framework of dissipative MHD taking into account 
viscosity, electrical and thermal conductivity, and radiative cooling of the plasma.
A piecewise homogeneous current layer model is used.
The simplicity of the model allows one to obtain accurate analytical expressions 
for the growth rate (Equation \ref{12}) and the spatial scale (Equation \ref{32}) of instability, 
as well as their simple approximations (Equations \ref{33}\,--\,\ref{37}) in the conditions of the solar corona.
The instability has a thermal nature. 
It occurs as a result of a drop in thermal conductivity inside the current layer 
and increases on the characteristic time scale of radiative plasma cooling. 
Due to the structural features of 
the radiative loss function 
of an optically thin medium, 
the spatial instability period is contained in a narrow range of values of about $l=1-10 {\rm \, Mm}$ 
for a wide range of parameters of the current layer and the surrounding plasma.

\begin{figure}
\begin{center}
\includegraphics*[width=0.5\linewidth]{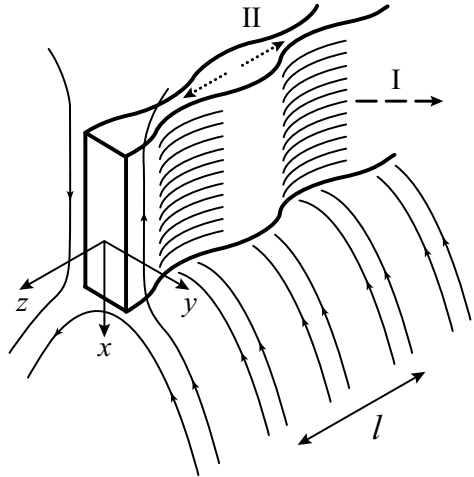}
\end{center}
\caption{Location of the perturbed current layer above the arcade of coronal magnetic loops. 
The perturbation has a spatial period $l$. 
The Roman numbers mark the energy fluxes associated with radiative plasma cooling (I) 
and heat conduction (II).}
\label{fig5}
\end{figure}
The instability properties allow us to offer the following qualitative picture of the solar flare triggering. 
There is a preflare current layer above the arcade of coronal magnetic loops (Figure \ref{fig5}). 
Due to a random perturbation, some of its sections begin to lose more heat by radiation. 
High electronic thermal conductivity can redistribute heat between cold and hot areas. 
However, if the electronic thermal conductivity is suppressed 
by the perturbation of the transverse magnetic field penetrating in the current layer, 
then the ionic thermal conductivity does not have time to transfer heat from hot to cold areas. 
The temperature difference between the cold and hot sections 
of the preflare current layer increases with increment described by Equation \ref{33}.
The alternation of cold and hot sections leads to a wave-like curvature 
of the surface of the current layer with a spatial period $l$
due to the total pressure balance.
The curvature has a symmetrical shape in accordance with the solution found.
The current layer begins to disintegrate into individual fibers 
located across the direction of the current, 
which can lead to its breaking and, as a result, to a solar flare. 
The regions of the main energy release will alternate with the same spatial period $l$. 
Flows of accelerated charged particles rush into the coronal magnetic loops 
located near the regions of energy release, which ultimately leads 
to the observed brightening of individual flare loops in the ultraviolet range.

In order to mathematically simplify the model, 
many significant physical features of the preflare current layer were neglected. 
Magnetic non-neutrality of the current layer leads to 
a change in the pressure balance at its boundary, while 
the appearance of a component of the magnetic field 
normal to the layer changes the type of MHD discontinuity on the boundary
\citep{1985SoPh...95..141S, 1985SoPh..102...79S}. 
The finite width of the current layer requires 
taking into account the corresponding derivatives with respect to the $ x $ coordinate, 
which leads to the appearance of tearing instability 
\citep{1988SoPh..117...89S, 1989SoPh..120...93S}. 
The observations of flare loops on the Sun indirectly indicate a complex current layer geometry 
that is different from a simple planar configuration. 
A statistical analysis of the flare loops themselves in the context of the considered model 
is a separate complex task. 
Attention on these and other issues will be paid in following articles of this series 
(Thermal Trigger for Solar Flares).

\begin{acks}
The author thanks Prof. Boris Somov, Vasilisa Nikiforova, and anonymous reviewer for discussing the article.

\noindent{\bf Disclosure of Potential Conflicts of Interest} The author declares that there are no conflicts of interest.
\end{acks}

\bibliographystyle{spr-mp-sola}
\bibliography{sola_bibliography}  

\end{article} 

\end{document}